# Complex-Dynamical Extension of the Fractal Paradigm and Its Applications in Life Sciences[*]

A.P. Kirilyuk

Institute of Metal Physics, 36 Vernadsky Av, Kiev-142, 03142 Ukraine
e-mail: kiril@metfiz.freenet.kiev.ua

**Summary.** Complex-dynamical fractal is a hierarchy of permanently, chaotically changing versions of system structure, obtained as the unreduced, causally probabilistic general solution to an arbitrary interaction problem. Intrinsic creativity of this extension of usual fractality determines its exponentially high operation efficiency, which underlies many specific functions of living systems, such as autonomous adaptability, "purposeful" development, intelligence and consciousness (at higher complexity levels). We outline in more detail genetic applications of complex-dynamic fractality, demonstrate the dominating role of genome interactions, and show that further progressive development of genetic research, as well as other life-science applications, should be based on the dynamically fractal structure analysis of interaction processes involved. We finally summarise the obtained extension of mathematical concepts and approaches closely related to their biological applications.

## 1   Introduction

The success of fractal paradigm in bio-system structure analysis, as presented in this series of conferences [1-3], reflects high efficiency of fractal geometry in life function realisation conceived and used by nature itself. In a broader sense, fractal structure efficiency appears inevitably and naturally in a wide variety of real processes, from physico-chemical structures to economic system evolution [4-8], driven by unreduced interaction processes and often referred to as systems with complex dynamics. Using the *universally nonperturbative* analysis of a generic interaction process, we have rigorously specified the connection between fractality and dynamic complexity [9,10], where the extended, complex-dynamic fractality has been derived as inevitably emerging structure of any real interaction process. In that way, the dynamic complexity as such acquires a rigorous and universally applicable definition, while the fractal structure of a real interaction is obtained as the truly complete, dynamically multivalued (probabilistic) *general solution* of a problem, replacing its reduced, dynamically single-valued (regular) version. The dynamically probabilistic, permanently changing fractal of real system dynamics is a natural extension of the canonical, "geometric" fractality possessing an involved, but basically predictable (regular) and fixed structure. Complex-dynamic fractality is not a "model" any more, but the *unreduced* version of any real, "nonintegrable" and "nonseparable" system structure and dynamics, which is especially interesting for fractality involvement with living systems because it provides rigorously derived versions of those essential life properties — such as intrinsic adaptability, self-development and "reasonable" behaviour — that determine its specific efficiency and remain largely "mysterious" within usual, perturbative theory.

   In this report, after recalling the mathematical framework of complex-dynamic fractality (section 2), we proceed to further exploration of its properties important for life-science applications. We show that due to the hierarchy of unceasing probabilistic change of the living fractal structure, its power to perform useful functions grows exponentially with the number of elements, contrary to power-law dependence in usual, dynamically single-valued models (section 3). Being applied to various important cases of interaction development in living organisms, such as genome dynamics or brain operation, this result explains their huge, qualitative advantages with respect to any conventional simulation that underlie all the "miracles of life" (self-reproduction, adaptable evolution, intelligence, consciousness, etc.). Important practical conclusions for genetic research strategy are derived from the unreduced fractal structure of genome interaction dynamics (section 4). In that way we substantiate and specify the necessary change in life sciences and related fields, which can uniquely solve the growing "difficult" (e.g. "ethical") problems of the modern blind, purely empirical technology development and provide the basis for the truly sustainable future. The latter involves genuine, *causally complete understanding* and control of living form emergence and dynamics, at any level of interest, giving rise to new possibilities in both fundamental (e.g. mathematical) and applied aspects of knowledge, including such directions as *constructive genetics* and *integral medicine* [9,10].

---
[*] Report presented at the IV[th] International Symposium "Fractals in Biology and Medicine" (Ascona, 10-13 March 2004), http://www.fractals.issi.cerfim.ch/.

## 2     Probabilistic fractal structure of a generic interaction process

We start from interaction problem between arbitrary (but known) system components, such as brain neurons, cell elements, or genes. It can be expressed by the *existence equation* that generalises many particular, model dynamic equations [9-13]:

$$\left\{ \sum_{k=0}^{N} \left[ h_k(q_k) + \sum_{l>k}^{N} V_{kl}(q_k, q_l) \right] \right\} \Psi(Q) = E\Psi(Q), \qquad (1)$$

where $h_k(q_k)$ is the "generalised Hamiltonian" of the *k*-th component in the absence of interaction with the degrees of freedom $q_k$, $V_{kl}(q_k, q_l)$ is the "interaction potential" between the *k*-th and *l*-th components, $\Psi(Q)$ is the system state-function depending on all degrees of freedom, $Q \equiv \{q_0, q_1, ..., q_N\}$, $E$ is the generalised Hamiltonian eigenvalue, and summations are performed over all (*N*) system components. The "Hamiltonian" equation form does not involve any real limitation and can be rigorously *derived*, in a self-consistent way, as a universal expression of real system dynamics [9,11,12], where generalised Hamiltonians express suitable measures of complexity defined below. One can present eq. (1) in another form, where one of the degrees of freedom, for example $q_0 \equiv \xi$, is separated because it represents an extended, common system component or measure (such as position of other, localised degrees of freedom and components):

$$\left\{ h_0(\xi) + \sum_{k=1}^{N} [h_k(q_k) + V_{0k}(\xi, q_k)] + \sum_{l>k}^{N} V_{kl}(q_k, q_l) \right\} \Psi(\xi, Q) = E\Psi(\xi, Q), \qquad (2)$$

where from now on $Q \equiv \{q_1, ..., q_N\}$ and $k,l \geq 1$.

The most suitable problem expression is obtained in terms of eigenfunctions $\{\varphi_{kn_k}(q_k)\}$ and eigenvalues $\{\varepsilon_{n_k}\}$ of non-interacting components:

$$h_k(q_k)\varphi_{kn_k}(q_k) = \varepsilon_{n_k}\varphi_{kn_k}(q_k), \qquad (3)$$

$$\Psi(\xi,Q) = \sum_{n \equiv (n_1, n_2, ..., n_N)} \psi_n(q_0)\varphi_{1n_1}(q_1)\varphi_{2n_2}(q_2)...\varphi_{Nn_N}(q_N) \equiv \sum_n \psi_n(\xi)\Phi_n(Q), \qquad (4)$$

where $\Phi_n(Q) \equiv \varphi_{1n_1}(q_1)\varphi_{2n_2}(q_2)...\varphi_{Nn_N}(q_N)$ and $n \equiv (n_1, n_2, ..., n_N)$ runs through all possible eigenstate combinations. Inserting eq. (4) into eq. (2) and performing the standard eigenfunction separation (e.g. by taking a scalar product), we obtain the system of equations for $\psi_n(\xi)$, which is equivalent to the starting existence equation:

$$[h_0(\xi) + V_{00}(\xi)]\psi_0(\xi) + \sum_n V_{0n}(\xi)\psi_n(\xi) = \eta\psi_0(\xi), \qquad (5a)$$

$$[h_0(\xi) + V_{nn}(\xi)]\psi_n(\xi) + \sum_{n' \neq n} V_{nn'}(\xi)\psi_{n'}(\xi) = \eta_n\psi_n(\xi) - V_{n0}(\xi)\psi_0(\xi), \qquad (5b)$$

where $n, n' \neq 0$ (also everywhere below), $\eta \equiv \eta_0 = E - \varepsilon_0$,

$$\eta_n \equiv E - \varepsilon_n, \quad \varepsilon_n \equiv \sum_k \varepsilon_{n_k}, \quad V_{nn'}(\xi) = \sum_k \left[V_{k0}^{nn'}(\xi) + \sum_{l>k} V_{kl}^{nn'}\right], \qquad (6)$$

$$V_{k0}^{nn'}(\xi) = \int_{\Omega_Q} dQ \Phi_n^*(Q) V_{k0}(q_k, \xi) \Phi_{n'}(Q), \quad V_{kl}^{nn'}(\xi) = \int_{\Omega_Q} dQ \Phi_n^*(Q) V_{kl}(q_k, q_l) \Phi_{n'}(Q), \qquad (7)$$

and we have separated the equation for $\psi_0(\xi)$ describing the generalised "ground state" of system elements, i. e. the state with minimum energy and complexity.

Now we try to "solve" eqs. (5) by expressing $\psi_n(\xi)$ through $\psi_0(\xi)$ from eqs. (5b) with the help of the standard Green function and substituting the result into eq. (5a), which gives the *effective existence equation* for $\psi_0(\xi)$ [9-13]:

$$h_0(\xi)\psi_0(\xi) + V_{\text{eff}}(\xi;\eta)\psi_0(\xi) = \eta\psi_0(\xi), \qquad (8)$$

where the *effective (interaction) potential (EP)*, $V_{\text{eff}}(\xi;\eta)$, is obtained as

$$V_{\text{eff}}(\xi;\eta) = V_{00}(\xi) + \hat{V}(\xi;\eta), \quad \hat{V}(\xi;\eta)\psi_0(\xi) = \int_{\Omega_\xi} d\xi' V(\xi,\xi';\eta)\psi_0(\xi'), \qquad (9a)$$

$$V(\xi,\xi';\eta) = \sum_{n,i} \frac{V_{0n}(\xi)\psi_{ni}^0(\xi)V_{n0}(\xi')\psi_{ni}^{0*}(\xi')}{\eta - \eta_{ni}^0 - \varepsilon_{n0}}, \quad \varepsilon_{n0} = \varepsilon_n - \varepsilon_0, \qquad (9b)$$

and $\{\psi_{ni}^0(\xi)\}$, $\{\eta_{ni}^0\}$ are complete sets of eigenfunctions and eigenvalues, respectively, for a truncated system of equations obtained as "homogeneous" parts of eqs. (5b):

$$[h_0(\xi) + V_{nn}(\xi)]\psi_n(\xi) + \sum_{n' \neq n} V_{nn'}(\xi)\psi_{n'}(\xi) = \eta_n\psi_n(\xi) . \qquad (10)$$

The eigenfunctions $\{\psi_{0i}(\xi)\}$ and eigenvalues $\{\eta_i\}$ found from eq. (8) are used to obtain other state-function components:

$$\psi_{ni}(\xi) = \int_{\Omega_\xi} d\xi' g_{ni}(\xi,\xi')\psi_{0i}(\xi'), \quad g_{ni}(\xi,\xi') = V_{n0}(\xi') \sum_{i'} \frac{\psi_{ni'}^0(\xi)\psi_{ni'}^{0*}(\xi')}{\eta_i - \eta_{ni'}^0 - \varepsilon_{n0}}, \qquad (11)$$

after which the total system state-function $\Psi(\xi,Q)$, eq. (4), is obtained as

$$\Psi(\xi,Q) = \sum_i c_i \left[\Phi_0(Q)\psi_{0i}(\xi) + \sum_n \Phi_n(Q)\psi_{ni}(\xi)\right], \qquad (12)$$

where coefficients $c_i$ should be found by state-function matching at the boundary where effective interaction vanishes. The observed (generalised) density, $\rho(\xi,Q)$, is obtained as state-function squared modulus, $\rho(\xi,Q) = |\Psi(\xi,Q)|^2$ (for "wave-like" complexity levels), or as state-function itself, $\rho(\xi,Q) = \Psi(\xi,Q)$ (for "particle-like" levels) [9].

Although the EP expression of a problem, eqs. (8)-(12), is formally equivalent to its initial version, eqs. (1), (2), (5), only the former reveals, due to its "dynamically rich" structure, the essential features designated as *dynamic multivaluedness (or redundance) and entanglement* and remaining hidden in the conventional formalism and especially its perturbative form of "exact" (or closed) solutions. Dynamic multivaluedness appears as redundant number of locally complete, and therefore *incompatible*, but *equally real* problem solutions, called *realisations*, while dynamic entanglement describes the related "cohesion" between interacting components within each realisation, expressing system "nonseparability". Because of equal reality and incompatibility of realisations, the system is forced, by the driving interaction itself, to permanently change them in a causally random order, forming each time a new version of component entanglement. The total number of eigen-solutions can be estimated by the maximum power of the characteristic equation for eq. (8). If $N_\xi$ and $N_Q$ are the numbers of terms in the sums over $i$ and $n$ in eq. (9b), equal to the numbers of system components ($N$) and their internal states, then the eigenvalue number is $N_{\max} = N_\xi(N_\xi N_Q + 1) = (N_\xi)^2 N_Q + N_\xi$, which gives the $N_\xi$-fold redundance of usual "complete" set of $N_\xi N_Q$ eigen-solutions of eqs. (5) plus an additional, "incomplete" set of $N_\xi$ eigen-solutions. The number of "regular" realisations is $N_\Re = N_\xi = N$, whereas the truncated set of solutions forms a specific, "intermediate" realisation that plays the role of transitional state during chaotic system jumps between "regular" realisations and provides thus the universal, *causally complete* extension of the quantum *wavefunction* and classical *distribution function* [9-13]. Note that dynamic multivaluedness is obtained only in the unreduced EP version (starting from the genuine quantum chaos description [14,15]), whereas practically all scholar applications of this well-known approach (see e.g. [16]) resort to its perturbative reduction that kills inevitably all manifestations of complex (chaotic) dynamics and is equivalent to the *dynamically single-valued, effectively zero-dimensional* (point-like) model of reality, containing only one, "averaged" system realisation (or projection).

The discovered multivaluedness of the unreduced solution and the ensuing chaoticity of *unceasing* realisation change are expressed by the *truly complete general solution* of a problem presenting the observed density $\rho(\xi,Q)$ (or a similar quantity) as the *causally probabilistic sum* of individual realisation densities, $\{\rho_r(\xi,Q)\}$:

$$\rho(\xi,Q) = \sum_{r=1}^{N_\Re} {}^\oplus \rho_r(\xi,Q), \qquad (13)$$

where summation over $r$ includes all observed realisations, while the sign $\oplus$ designates the causally probabilistic sum. The dynamically probabilistic general solution of eq. (13) is accompanied by the *dynamically derived* values of *realisation probabilities* $\alpha_r$:

$$\alpha_r = \frac{N_r}{N_\Re} \ \left( N_r = 1,\ldots,N_\Re;\ \sum_r N_r = N_\Re \right), \quad \sum_r \alpha_r = 1, \qquad (14)$$

where $N_r$ is the number of elementary realisations grouped in the $r$-th "compound" realisation, but remaining unresolved in a general case. It is important that eqs. (13), (14) contain not only the ordinary "expectation" value for a large series of events, but remain also valid for any *single* event observation and even before it, providing *a priori probability* and its *universal dynamic origin*. A practically useful probability definition is given also by the generalised Born rule [9,11,12], derived by dynamic matching and presenting the wavefunction in a physically transparent form of *probability distribution density* (or its amplitude, for the "wave-like" levels of complexity):

$$\alpha_r = |\Psi(X_r)|^2 ,\qquad(15)$$

where $X_r$ is the *r*-th realisation configuration, while the wavefunction can be found from the *universal, causally derived Schrödinger equation* [9,11,12].

*Dynamic complexity*, $C$, can be *universally* defined now as any growing function of system realisation number, or rate of their change, equal to zero for only one system realisation: $C = C(N_\Re)$, $dC/dN_\Re > 0$, $C(1) = 0$. It is the latter case of zero unreduced complexity that is invariably considered in the canonical, dynamically single-valued, or *unitary*, theory, which explains all its old and new difficulties at various levels of world dynamics [9-13]. The unreduced dynamic complexity is presented by the majority of actually measured quantities, such as energy, mass, momentum, action, and entropy, now provided with a universal and *essentially nonlinear* interpretation in terms of the underlying interaction processes. *Space* and *time* are two universal, physically real *forms of complexity*, causally derived as *tangible quality* of *dynamically entangled* structure and *immaterial* rate (frequency) of realisation change *events*, respectively. Complex dynamics is a structure emergence process (*dynamically multivalued self-organisation*) and can be described by the *universal Hamilton-Jacobi equation* for the generalised action, which is dualistically related to the universal Schrödinger equation mentioned above through the *causal quantization condition* (it reflects realisation change by transition through the intermediate realisation of the wavefunction) [9,11,12]. Note finally that dynamic complexity thus defined represents at the same time universal measure of *genuine* and omnipresent *chaoticity* and *(generalised) entropy*.

The *complex-dynamic, intrinsically probabilistic fractality* represents the inevitable development and internal content of dynamic entanglement (nonseparability), complexity and chaoticity. It is related to problem *nonintegrability* as it appears in EP dependence on the unknown solutions of the auxiliary system of equations, eqs. (10). After we have revealed dynamic system splitting into chaotically changing realisations at the first level of nonperturbative dynamics, we should now proceed with further analysis of the auxiliary system solutions, which introduce additional structure in the general solution. Due to the unrestricted universality of the generalised EP method, it can be applied to the truncated system (10), transforming it into a single effective equation, quite similar to the first-level EP result of eq. (8):

$$\left[h_0(\xi) + V_{\text{eff}}^n(\xi;\eta_n)\right]\psi_n(\xi) = \eta_n \psi_n(\xi),\qquad(16)$$

where the second-level EP action is analogous to the combined version of eqs. (9):

$$V_{\text{eff}}^n(\xi;\eta_n)\psi_n(\xi) = V_{nn}(\xi)\psi_n(\xi) + \sum_{n'\neq n, i} \frac{V_{nn'}(\xi)\psi_{n'i}^{0n}(\xi)\int_{\Omega_\xi} d\xi'\psi_{n'i}^{0n*}(\xi')V_{n'n}(\xi')\psi_n(\xi')}{\eta_n - \eta_{n'i}^{0n} + \varepsilon_{n0} - \varepsilon_{n'0}},$$

$$(17)$$

and $\{\psi_{n'i}^{0n}(\xi), \eta_{n'i}^{0n}\}$ is the eigen-solution set for the second-level truncated system:

$$h_0(\xi)\psi_{n'}(\xi) + \sum_{n'' \neq n'} V_{n'n''}(\xi)\psi_{n''}(\xi) = \eta_{n'}\psi_{n'}(\xi), \quad n' \neq n, \quad n,n' \neq 0. \tag{18}$$

The same mechanism of dynamic multivaluedness due to the *essentially nonlinear* EP dependence on the eigen-solutions to be found in eqs. (16)-(17) leads to the second level of splitting, this time of auxiliary system solutions entering the first-level expressions (8)-(12), into many mutually incompatible realisations (numbered by index $r'$):

$$\{\psi_{ni}^0(\xi), \eta_{ni}^0\} \rightarrow \{\psi_{ni}^{0r'}(\xi), \eta_{ni}^{0r'}\}. \tag{19}$$

We can continue to trace this hierarchy of dynamical splitting by applying the same EP method to ever more truncated systems of equations, such as eqs. (18), and obtaining corresponding levels of dynamically multivalued structures with the attached intrinsic space and time, until we obtain a directly integrable equation for one unknown function. The maximum number of levels in this dynamically multivalued hierarchy is equal to the number of component states (excitations) $N_Q$, although in practice each of them need not be resolved. We can now specify the detailed, *probabilistically fractal* structure of the complete general solution to the interaction problem, eq. (13):

$$\rho(\xi,Q) = \sum_{r,r',r''...}^{N_\Re} {}^\oplus \rho_{rr'r''...}(\xi,Q), \tag{20}$$

with indexes $r,r',r''$,... enumerating permanently, chaotically changing realisations of consecutive levels of *dynamic (probabilistic) fractality*, naturally emerging thus as the *unreduced, truly exact solution* to any real many-body problem, eqs. (1), (2), (5). The time-averaged *expectation value* for the dynamically fractal density is given by

$$\rho(\xi,Q) = \sum_{r,r',r''...}^{N_\Re} \alpha_{rr'r''...} \rho_{rr'r''...}(\xi,Q), \tag{21}$$

where the *dynamically determined probabilities* of the respective fractality levels are obtained in a form analogous to eq. (14)

$$\alpha_{rr'r''...} = \frac{N_{rr'r''...}}{N_\Re}, \quad \sum_{rr'r''...} \alpha_{rr'r''...} = 1. \tag{22}$$

Multivalued fractal solution of eqs. (20)-(22) can be obtained in a number of versions, but with the same essential result of probabilistically adapting hierarchy of realisations. Consecutive level emergence of unreduced dynamic fractality should be distinguished from perturbative series expansion: the latter provides a *qualitatively incorrect*, generically "diverging" (*because* of dynamic single-valuedness [9]) *approximation* for a *single* level of structure, while the series of levels of dynamic fractality corresponds to *really emerging* structures, where each level is obtained in its unreduced, dynamically multivalued and entangled version. In fact, the ultimately complete, dynamically fractal version of the general solution demonstrates the genuine, physically transparent origin of a generic problem "nonintegrability" (absence of a "closed", unitary solution) and related "nonseparability" (now being clearly due to the physical, fractally structured and chaotically changing component entanglement).

The dynamically probabilistic fractal thus obtained is a natural extension of the ordinary, dynamically single-valued (basically regular) fractality, which is especially important for life-science applications because it possesses the essential living system properties absent in any unitary model, including autonomous dynamic adaptability, "purposeful" self-development, intrinsic mixture of omnipresent randomness with often implicit but strong order, and the resulting qualitatively superior dynamic efficiency. These properties are unified within the universal dynamic *symmetry, or conservation, of complexity* [9,11,12] providing the general framework for the described process of interaction development into a probabilistically fractal structure. The initial interaction configuration, as described by the starting equations (1), (2), (5), is characterised by the latent, "potential" complexity form of *dynamic information*, universally measured by generalised action. System structure emergence in the form of unreduced dynamical fractal, eqs. (8)-(22), is described by unceasing transformation of dynamic information into a dual complexity form, *dynamic entropy*, generalising the usual entropy to any real system dynamics and reflecting the fully developed structure. Symmetry of complexity means that the sum of dynamic information and entropy, or *total complexity*, remains *unchanged* for any given system or process, which gives rise to the universal Hamilton-Schrödinger formalism mentioned above and extended, causally complete versions of all other (correct) laws and principles. Due to the intrinsic randomness of the unreduced fractality and contrary to any unitary symmetry, the universal symmetry of complexity relates *irregular*, configurationally "asymmetric" structures and elements, while remaining always *exact* (unbroken), which is especially important for description of biological, explicitly irregular, but *internally ordered* structures. Constituting thus the unreduced symmetry of natural structures, the symmetry of complexity extends somewhat too regular symmetry of usual fractals and approaches the fractal paradigm to the unreduced complexity of living organism structure and dynamics.

## 3   Exponentially high efficiency of unreduced fractal dynamics

The probabilistic dynamical fractal, eqs. (8)-(22), emerges as a *single whole*, which means that the fractal hierarchy of realisations appears and adapts its structure in a "real-time" period, comparable with the time of structure formation of the first level of fractality. This is the *complex-dynamical, multivalued*, genuine *parallelism* of real system dynamics absent in unitary models that try to imitate it by *artificial* division of *sequential* thread of events between simultaneously working *multiple units* of interaction, which can be useful, but does not provide any true gain in power. By contrast, the real, exponential power increase is obtained in natural systems with many interacting units at the expense of *irreducible dynamic randomness*, which constitutes the necessary, but actually quite advantageous "payment" for the huge power growth of *creative* interaction processes (whereas any unitary, regular dynamics is strictly deprived of genuine creativity).

System operation power $P$ is proportional to the number of realisations emerging within a given time interval, i.e. to the unreduced dynamic complexity: $P = P_0 C(N_\Re)$, where $P_0$ is a coefficient conveniently taken to be equal to the corresponding unitary power value (dynamically single-valued, sequential operation model, or "generalised Turing machine"). Then the relative growth of complex-dynamical fractal power with respect to unitary model, $\delta P$, is given by the unreduced system complexity, which can be estimated by the fractal realisation number: $\delta P = P/P_0 = C(N_\Re) = N_\Re - 1 \cong N_\Re$ $(N_\Re \gg 1)$. According to the analysis of section 2, we have the complex-dynamical fractal hierarchy of system realisations with $N_Q$ levels, each of them producing a new split into $N_\xi = N$ realisations (where $N$ is the number of system components and $N_Q$ is the number of their operative states). So the total (maximum) realisation number $N_\Re$ of the dynamical fractal, and thus also $\delta P$, grows exponentially with $N_Q$:

$$\delta P \cong N_\Re = N^{N_Q}. \tag{23}$$

Truly complicated systems from superior complexity levels, such as genome, cell, or brain dynamics, have high values of $N$ and $N_Q$, so that their exponential combination of eq. (23) produces not only quantitative, but also *qualitative* effects appearing as various "miracles" of "living" and "intelligent" behaviour that cannot be convincingly imitated by unitary models (and now we know the *exact, fundamental* reason for that).

The estimate of eq. (23) refers, however, to a single interaction "run" at a given level of complexity describing the emergence of one "compound", fractally structured realisation of the first level. System structure formation in the process of its operation does not stop there and involves a hierarchy of interactions at superior levels, where the above fractal structure within a given level plays the role of distributed "interaction transmitter" between harder, first-level parts of fractality. This means that the dynamic fractal grows, starting from a given interaction level, not only "in depth" (to generally smaller scales and lower complexity sublevels), but also to higher complexity levels. In order to estimate the total relative efficiency of such systems of "biologically high" complexity, consider a many-body interaction system consisting of $N_\text{unit}$ operative units (such as neurons, or genes, or relevant cell components) each of them connected by $n_\text{link}$ effective links to other units, so that the total number of interaction links in the system is $N = N_\text{unit} n_\text{link}$. The number of system realisations $N_\Re$, and thus $\delta P$, is of the order of the number of *all possible combinations of links*, $N_\Re \simeq N!$, which is the distinctive feature of the unreduced, *dynamically multivalued* fractality [11]:

$$\delta P = N_\Re \simeq N! \simeq \sqrt{2\pi N}\left(\frac{N}{e}\right)^N \sim N^N, \tag{24}$$

where we have used the well-known Stirling formula valid for large $N$ (which is greater than $10^{12}$ for both brain and genome interaction structure, see section 4). For the case of $N \sim 10^{12}$ the estimate of eq. (24) gives $\delta P \gg 10^{10^{13}} \gg 10^{10^{12}} \sim 10^N$, which is a practical infinity demonstrating the *qualitatively* huge efficiency of complex-dynamic fractality and its causal origin. Note that any unitary (basically regular and sequential) model of the same system dynamics would give the operation power growing only as $N^\beta$ ($\beta \sim 1$) and remaining negligible with respect to exponentially big efficiency of unreduced complex dynamics (including its unique adaptability and creativity).

## 4    Causally complete genetics, integral medicine, and other applications of the unreduced complex-dynamic fractality

Causally complete understanding of complex-dynamical fractal structure development in real biological and bio-inspired systems leads to a number of promising applications in life sciences, where modification and control of bio-system dynamics deal with its realistic, unreduced version and are comparable with natural creation processes. The relevant examples include (see also [9-11]) (1) causally complete understanding and use of the natural *biological evolution dynamics*, involving both relative permanence and sudden "reasonable" change of species; (2) *causally complete genetics* taking into account the whole picture of real genome interactions and thus providing the desirable and reliable modifications; (3) unreduced understanding of the *brain dynamics* and *emergent*, dynamic properties of *intelligence and consciousness*; (4) *integral medicine* based on the causally complete understanding and creative control of each individual organism dynamics; (5) genuine paradigm of *nanotechnology* based on the *irreducibly complex (multivalued) dynamics of nano-scale structures* approaching them to the natural, biological nano-machines; and (6) *ecological* and *social* applications of the unreduced (multivalued) fractality and complexity characterised by the intrinsically *holistic* analysis of the multi-level systems involved and providing *provably efficient* solutions to the "global" problems (that *cannot* be solved within the unitary approach, irrespective of the quantity of efforts [9]). Only such *unreduced understanding of real system dynamics* can solve the growing "ethical" problems in practical research.

We shall consider here a more detailed outline of genetic applications, as they become especially important because of the growing conceptually blind, but technically powerful empirical experimentation with genomes of various organisms. The key result, strongly supported by both experimental knowledge and the above theory, is that the genome structure, operation, evolution, and related organism phenotype are *mainly* determined by fractally structured *genome interactions* and *not* by sequential "programme reading" à la Turing machine, as it is assumed by the current theory and applications. Such understanding of genome dynamics is supported by the ensuing unified solution to the well-known problem of "noncoding DNA", relatively large in quantity, but apparently "useless", in the framework of unitary genetic paradigm. We can see now that the existence of those relatively large DNA sections is *necessary* as *fractally structured gene interaction space and transmitter*, similar to any real interaction process and in agreement with experimentally observed correlation between organism complexity and relative volume of those noncoding DNA parts [17].

As follows from sections 2 and 3, a unitary genetic programme cannot provide "reasonable" development and would actually halt in any realistic operation mode. Its efficiency is smaller than that of a real, dynamically multivalued, fractal interaction process by a practically infinite quantity given by eq. (24). Unfortunately, this does not exclude a possibility of purely empirical genome modification whose *immediate* consequences, considered only within severely reduced unitary model, cover only a *negligibly small part* of actually introduced change in the *whole system dynamics*, remaining delayed in time and therefore "hidden" in mechanistic experimentation.

As has been shown in section 3, the huge dynamic complexity of brain or genome operation is determined by the number of links between the system elements. The number of synaptic links in human brain can be estimated as $N_{\text{brain}} = N_{\text{neuron}} n_{\text{syn}} \approx 10^{10} \times 10^4 = 10^{14}$, where $N_{\text{neuron}} \approx 10^{10}$ is the number of cells and $n_{\text{syn}} \approx 10^4$ is the number of links per cell. As follows from the universal symmetry of complexity (section 2), the number of interaction links in the genome $N_{\text{genome}}$, determining the emerging brain complexity, cannot be smaller than $N_{\text{brain}}$, $N_{\text{genome}} \geq N_{\text{brain}}$. Since $N_{\text{genome}} = N_{\text{gene}} n_{\text{eff}}$, where $N_{\text{gene}}$ is the number of genes and $n_{\text{eff}}$ is the number of interaction links per gene, we have $n_{\text{eff}} \geq N_{\text{brain}}/N_{\text{gene}} \approx 3 \times 10^9$ for human genome ($N_{\text{gene}} \approx 3 \times 10^4$). It is remarkable that not only $n_{\text{eff}}$ is quite large, supporting the key role of gene interaction (both direct and indirect one), but in fact $n_{\text{eff}} \geq N_{\text{base}}$, where $N_{\text{base}} \approx 3 \times 10^9$ is the experimentally determined number of smallest chemical elements ("bases") in the human genome. This strongly supports the above idea that the main part of genome is playing the role of effective "interaction space" and only its smaller part appears as relatively "condensed", stable, coding gene sequences (also contributing to omnipresent interaction links through various transmitting agents). The fact that $N_{\text{genome}} \geq N_{\text{gene}} N_{\text{base}}$ shows that interactions of *each* (average) gene involve, in one way or another, *any individual base* and the reverse, any (average) base participates in every gene operation. Such incredible wholeness of the huge system of genome interactions can be realised only through the *probabilistic fractal hierarchy* of *emerging* system realisations, in agreement with the detailed analysis of sections 2, 3. It is interesting that for human genome and brain we have $N_{\text{brain}} = N_{\text{neuron}} n_{\text{syn}} \approx N_{\text{gene}} N_{\text{base}} \approx 10^{14}$, which confirms the symmetry of complexity [9-12] unifying the *probabilistically developing* fractal of human organism dynamics into a single whole, from genome information unfolding to the brain operation. We can apply the same, universal understanding of fractal interaction dynamics and its exponentially high efficiency to other biological and bio-inspired systems of particular interest today, such as neuron system dynamics and its "higher" properties known as intelligence and consciousness [9,11], various aspects of cell dynamics, artificial nanosystems [11], ecological and social systems, etc.

The probabilistically changing, fractal hierarchy of genome dynamics provides also the necessary combination of relative stability of a species genome and its capacity for rare evolutionary changes. The latter can now be causally understood as the largest, most "coarse-grained" level of probabilistic realisation change at the level of whole genome and organism dynamics. Such "global" changes are prepared by hidden potentialities accumulated from all interactions in the genome-organism-environment system and particularly "activated" in a "period of change" characterised by especially heavy pressure of the environment and critically dominating defects of genome dynamics. Those *real* potentialities for a future "big" change cannot appear as such before the change and remain *hidden* somewhere in the exponentially large, fractally involved space of genome interactions, thus ensuring the necessary (but always *limited*) *stability* of species genome in a period *between* those big, evolutionary changes. Therefore it becomes evident that empirically based artificial modifications of any organism genome (related by a fractal interaction network to other organisms) will produce absolutely unknown and unpredictable (but typically destructive) effect on higher-level interactions that will appear in their *explicitly observable* form *only* during the next period of "big" change, remaining until then *hidden* behind superficially smooth "everyday" level of organism dynamics. That the "big change" will come inevitably in an evolutionary short period of time follows from the same symmetry of complexity, which leads to the *causally substantiated* conclusion about the *fundamentally limited life cycle* of *any* system, including a biological species and its ecological niche. It is determined by the *complete transformation* of system interaction complexity from "potentialities" (dynamic information) to "reality" (dynamic entropy), where characteristic, observable signs of approaching "bifurcation" can be predicted [9] and correlate with a number of currently growing "criticality" features. The technically powerful, but conceptually blind genetic experimentation of today can be compared in this sense to charging of *delayed-action "genetic bomb", or G-bomb*, another potential weapons of *mass* destruction (though remaining unpredictable in details), where the "charging" process has a transparent physical meaning of introducing additional,

"unnatural" tensions in the "infinitely" large network of fractal interactions (eq. (24)), among which only some part will explicitly appear in the observed properties of organism dynamics. It is also evident that the problem can be solved *only* by *essential extension* of unitary approaches to the unreduced, multivalued and fractal interaction dynamics, taking into account all participating elements, as it is demonstrated by the above analysis, which can uniquely transform the empirical, potentially destructive unitary genetics into *provably constructive complex-dynamical genetics*.

    Note finally the essential extension of *mathematical concepts and approaches* involved with that urgently needed progress in applications, as the development of fundamental science tools represents also its own interest, especially evident on the background of persisting stagnation [9,11,18] and "loss of certainty" in fundamental knowledge (cf. [19]). (i) First of all, one should mention the *non*uniqueness of *any real* problem solution, taking the form of its *dynamic multivaluedness* (section 2), and related *complex-dynamic existence* of any system that replace the usual "uniqueness and existence theorems" valid only for reduced, unitary models [9]. (ii) It follows that the related unitary concept of "exact" (closed) solutions and its perturbative versions are basically insufficient and fundamentally incorrect with respect to real world structures. The true, dynamical meaning of the notions of "(non)integrability", "(non)separability", "(non)computability", "uncertainty", "randomness", and "probability" becomes clear: we obtain now the *nonintegrable and nonseparable, but solvable* dynamics of a generic many-body system (see eqs. (8)-(22)), while real world mathematics regains its *certainty* and *unification*, but contains a *well-defined, dynamic indeterminacy* and fractally structured *diversity* (i.e. it *cannot* be reduced to number properties and geometry, contrary to unitary hopes). (iii) The property of *dynamic entanglement* and its fractal extension (section 2) provides the *rigorous mathematical definition* of the tangible *quality* of a structure, applicable at any level of dynamics, which contributes to the *truly exact* mathematical representation of real objects, especially important for biological applications. (iv) The irreducible *dynamic discreteness, or quantization*, of real interaction dynamics expresses its *holistic* character and introduces essential modification in standard calculus applications and their *formally* discrete versions, including "evolution operators", "Lyapunov exponents", "path integrals", etc. [9,11]. (v) The unceasing, probabilistic *change* of system realisations provides the *dynamic origin of time*, absent in any version of unitary theory: in the new mathematics and in the real world one always has $a \neq a$ for any measurable, realistically expressed quantity or structure $a$, while one of the basic, often implicit postulates of the canonical mathematics is "self-identity", $a = a$ (related to "computability"). It has a direct bio-inspired implication: *every* real structure $a$ is "alive" and "noncomputable", in the sense that it always probabilistically moves and changes internally. In fact, *any* realistically conceived $a$ represents a part of a *single, unified structure* of the new mathematics introduced above as *dynamically multivalued (probabilistic) fractal* (of the world structure) and obtained as the *truly exact, unreduced solution* of a *real* interaction problem (section 2). We can see in that way that such recently invented terms as "biofractals" and "biomathematics" can have much deeper meaning and importance than usually implied "(extensive) use of mathematics in biological object studies".

**Acknowledgement**. The author is grateful to Professor Gabriele Losa for invitation to the symposium and support.